\documentclass[prb,twocolumn,preprintnumbers,superscriptaddress,amsmath,amssymb]{revtex4-2}
\usepackage[dvipsnames]{xcolor}
\usepackage{graphicx}
\usepackage{subfigure}
\usepackage{mathrsfs}
\usepackage{amsfonts}
\usepackage{times}
\usepackage{amsmath}
\usepackage{leftidx}
\usepackage[colorlinks,linkcolor=blue,citecolor=blue]{hyperref}

\usepackage{bbold}
\usepackage{braket}
\usepackage{mathtools}
\usepackage{colortbl}
\usepackage{multirow}

\newcommand{\bs}[1]{\boldsymbol{#1}}

\usepackage[normalem]{ulem}

\newcommand{\ii}{\mathrm{i}}

\newcommand{\ie}{{\it i.e.},\ }
\newcommand{\eg}{{\it e.g.},\ }
\newcommand{\bb}{{\rm b}}
\newcommand{\ff}{{\rm f}}
\begin{document}

\title{Supersymmetric free fermions and bosons: Locality, symmetry and topology}
\author{Zongping Gong}
\affiliation{Max-Planck-Institut f\"ur Quantenoptik, Hans-Kopfermann-Stra{\ss}e 1, 85748 Garching, Germany}
\affiliation{Munich Center for Quantum Science and Technology, Schellingstra{\ss}e 4, 80799 M\"unchen, Germany}
\author{Robert H. Jonsson}
\affiliation{Max-Planck-Institut f\"ur Quantenoptik, Hans-Kopfermann-Stra{\ss}e 1, 85748 Garching, Germany}
\affiliation{Munich Center for Quantum Science and Technology, Schellingstra{\ss}e 4, 80799 M\"unchen, Germany}
\author{Daniel Malz}
\affiliation{Max-Planck-Institut f\"ur Quantenoptik, Hans-Kopfermann-Stra{\ss}e 1, 85748 Garching, Germany}
\affiliation{Munich Center for Quantum Science and Technology, Schellingstra{\ss}e 4, 80799 M\"unchen, Germany}
\date{\today}

\begin{abstract}
Supersymmetry, originally proposed in particle physics, refers to a dual relation that connects fermionic and bosonic degrees of freedom in a system. Recently, there has been considerable interest in applying the idea of supersymmetry to topological phases, motivated by the attempt to gain insights from the fermion side into the boson side and vice versa. We present a systematic study of this construction
when applied to band topology in noninteracting systems. 
First, on top of the conventional ten-fold way, we find that topological insulators and superconductors are divided into three classes depending on whether the supercharge can be local and symmetric, must break a symmetry to preserve locality, or needs to break locality.
Second, we resolve the apparent paradox between the nontriviality of free fermions and the triviality of free bosons by noting that the topological information is encoded in the identification map. We also discuss how to understand a recently revealed supersymmetric entanglement duality in this context. These findings are illustrated by prototypical examples. Our work sheds new light on band topology from the perspective of supersymmetry.
\end{abstract}
\maketitle

\section{Introduction}
Supersymmetric (SUSY) models play an important role in physics. 
Perhaps the most well-known is their use in relativistic quantum field theories, where they resolve a number of theoretical problems~\cite{MirrorSymmetryBook}.
SUSY in quantum mechanics was introduced to understand properties of SUSY theories~\cite{Nicolai1976}, which led to profound understanding of SUSY breaking~\cite{Witten1981}.
SUSY can also appear in non-relativistic theories, for example, in statistical mechanics~\cite{Friedan1984}, or the Sachdev-Ye-Kitaev model~\cite{Behrends2020}. Besides, it is a powerful tool in the analysis of disordered systems~\cite{Efetov1996} or to solve an array of problems in quantum and statistical physics~\cite{Junker1996}.

In the last decades, topology has risen to become a cornerstone of our understanding of condensed matter physics~\cite{Kane2010,Qi2011,Ryu2016}.
For example, the well-known periodic table \cite{Ryu2008,Ryu2010,Kitaev2009,Teo2010} exhausts the topological classifications of insulators and superconductors in the  ten fundamental symmetry (Altland-Zirnbauer) classes \cite{Altland1997} in all dimensions.

Recently,  the idea of SUSY has been applied to extract the topological indices of free-boson systems, such as photonic and magnonic crystals, from their free-fermion counterparts \cite{Shindou2013,Lu2018,Roychowdhury2018,Attig2019}. This strategy indeed works well for individual bands, which may still be nontrivial despite the fact that the ground state of free bosons is always trivial \cite{Xu2020}. In seeming contradiction, one important measure of topology, the entanglement spectrum \cite{Fidkowski2010} has recently found to translate from the fermionic \emph{ground state} to the bosonic ground state~\cite{Jonsson2021}.

\begin{figure}
	\includegraphics[width=8.4cm, clip]{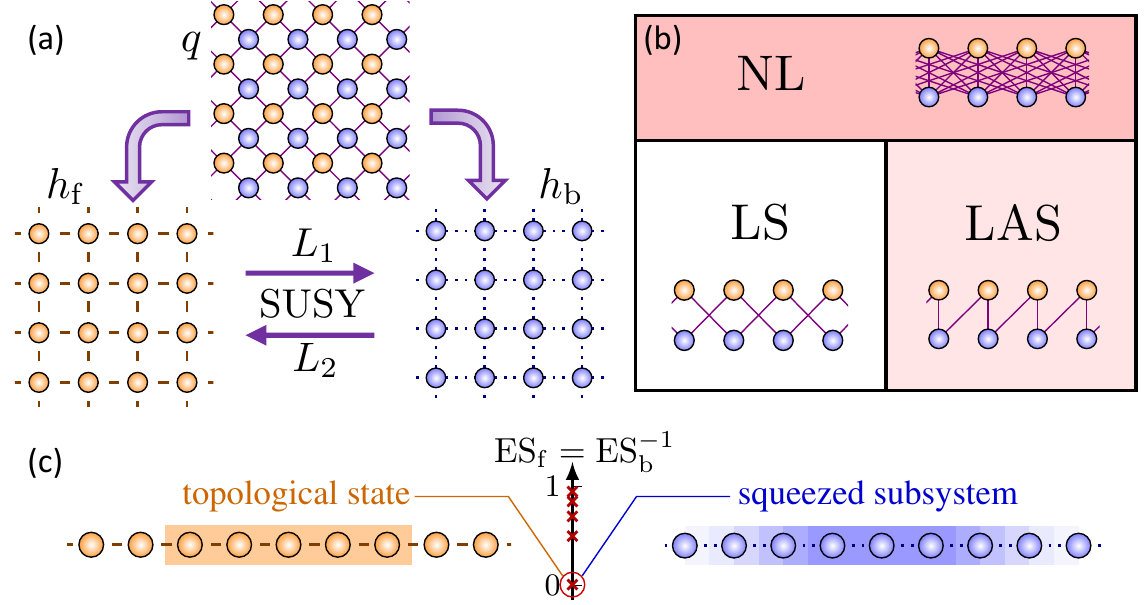}
	\caption{(a) A translation-invariant supercharge generates a pair of SUSY fermion and boson systems. The SUSY pair can be related to each other by the identification maps.  (b) Free-fermion topological phases can be categorized into three classes, depending on whether their parent supercharges (i) are necessarily non-local (NL); (ii) can be local but then necessarily asymmetric (LAS); (iii) can be both local and symmetric (LS) (see Table~\ref{table1}). (c) Entanglement-spectrum (ES) duality for two subsystems related by the identification map. For topological free-fermion phases, the subsystem on the bosonic side contains strongly squeezed modes.}
	\label{fig1}
\end{figure}

In this work, we analyze free-SUSY systems systematically. We start from a basic question concerning the existence of a SUSY construction for free-fermion topological phases. The answer turns out to be highly nontrivial: depending on the topological class, the SUSY generator, called  \emph{supercharge}, may be necessarily non-local, or local but necessarily asymmetric \footnote{More precisely speaking, the supercharge does not respect all the symmetries of the Hamiltonian.} [cf. Fig.~\ref{fig1}(b)]. The entanglement problem gets resolved as the SUSY map, called \emph{identification map}, can lead to very strong squeezing, if it is locality-preserving.

\section{Supersymmetric bands}
We consider a pair of fermion and boson systems in $d$ dimensions ($d$D). For simplicity,  we assume that fermions and bosons live on identical hypercubic lattices $\Omega\subset\mathbb{Z}^d$ with a set of internal states $I$ at each unit cell and subject to periodic boundary conditions. Denoting the fermionic~/~bosonic modes by $\hat c_{\boldsymbol{r}s}$~/~$\hat b_{\boldsymbol{r}s}$ ($\boldsymbol{r}\in\Omega$, $s\in I$), we can write down a general translation invariant supercharge
\begin{equation}
\hat Q= \sum_{\boldsymbol{k}} \begin{bmatrix} \hat{\boldsymbol{c}}^\dag_{\boldsymbol{k}} \\ \hat{\boldsymbol{c}}_{-\boldsymbol{k}} \end{bmatrix}^{\rm T} q(\boldsymbol{k})  \begin{bmatrix} \hat{\boldsymbol{b}}_{\boldsymbol{k}} \\ \hat{\boldsymbol{b}}^\dag_{-\boldsymbol{k}} \end{bmatrix},
\label{sck}
\end{equation}
where $\hat{\boldsymbol{c}}_{\boldsymbol{k}}\equiv[\hat c_{\boldsymbol{k}s}]^{\rm T}_{s\in I}$, $\hat{\boldsymbol{b}}_{\boldsymbol{k}}\equiv[\hat b_{\boldsymbol{k}s}]^{\rm T}_{s\in I}$ with $\hat c_{\boldsymbol{k}s}=|\Omega|^{-\frac{1}{2}}\sum_{\boldsymbol{r}\in\Omega} e^{-\ii\boldsymbol{k}\cdot\boldsymbol{r}} \hat c_{\boldsymbol{r}s}$, $\hat b_{\boldsymbol{k}s}=|\Omega|^{-\frac{1}{2}}\sum_{\boldsymbol{r}\in\Omega} e^{-\ii\boldsymbol{k}\cdot\boldsymbol{r}} \hat b_{\boldsymbol{r}s}$ ($|\cdot|$: cardinality of a set).
The $2|I|\times 2|I|$ matrix $q(\boldsymbol{k})$ satisfies $q(\boldsymbol{k}+2\pi \boldsymbol{e}_j)=q(\boldsymbol{k})$ ($\boldsymbol{e}_j$: unit vector in the $j$th direction) $\forall j=1,2,...,d$, and
\begin{equation}
Xq(\boldsymbol{k})^*X = q(-\boldsymbol{k}),\;\;\;\;X\equiv\sigma_x\otimes \openone.
\label{qPHS}
\end{equation}
Here, $\sigma_\mu$ ($\mu=x$) is the Pauli matrix and $\openone$ is the identity acting on the internal states. This symmetry (\ref{qPHS}) arises from the Hermiticity of the supercharge. If the fermion-boson coupling in $\hat Q$ is short-ranged, i.e., decays no slower than exponentially, we know that $q(\boldsymbol{k})$ is analytic in $\boldsymbol{k}$ and vice versa \cite{Ashida2021}. We will call such a supercharge local.

By taking the square of the supercharge (\ref{sck}), we obtain a SUSY pair of fermion and boson bands [cf. Fig.~\ref{fig1}(a)]
\begin{equation}
\begin{split}
&\hat H=  \hat Q^2 = \hat H_{\rm f} + \hat H_{\rm b}, \\
&\hat H_{\rm f} = \frac{1}{2}\sum_{\boldsymbol{k}} \begin{bmatrix} \hat{\boldsymbol{c}}^\dag_{\boldsymbol{k}} \\ \hat{\boldsymbol{c}}_{-\boldsymbol{k}} \end{bmatrix}^{\rm T} 
h_{\rm f}(\boldsymbol{k})
\begin{bmatrix} \hat{\boldsymbol{c}}_{\boldsymbol{k}} \\ \hat{\boldsymbol{c}}^\dag_{-\boldsymbol{k}} \end{bmatrix}, \\
&\hat H_{\rm b}= \frac{1}{2}\sum_{\boldsymbol{k}}\begin{bmatrix} \hat{\boldsymbol{b}}^\dag_{\boldsymbol{k}} \\ \hat{\boldsymbol{b}}_{-\boldsymbol{k}} \end{bmatrix}^{\rm T} 
h_{\rm b}(\boldsymbol{k})
\begin{bmatrix} \hat{\boldsymbol{b}}_{\boldsymbol{k}} \\ \hat{\boldsymbol{b}}^\dag_{-\boldsymbol{k}} \end{bmatrix},
\end{split}
\label{HfHb}
\end{equation}
where the Bogoliubov-de Gennes (BdG) Hamiltonians $h_{\rm f}(\boldsymbol{k})$ and $h_{\rm b}(\boldsymbol{k})$ are determined by $q(\boldsymbol{k})$ via
\begin{equation}
\begin{split}
&h_{\rm f}(\boldsymbol{k}) = q(\boldsymbol{k})Zq(\boldsymbol{k})^\dag,\;\;\;\;Z\equiv\sigma_z\otimes\openone, \\
&h_{\rm b}(\boldsymbol{k}) = q(\boldsymbol{k})^\dag q(\boldsymbol{k}).
\end{split}
\label{hfhb}
\end{equation}
The full derivation of the above results are given in Appendix~\ref{DSUSYH}. As an important implication of SUSY, the dynamical matrices $h_{\rm f}(\boldsymbol{k})$ and $Zh_{\rm b}(\boldsymbol{k})$ share exactly the same spectrum \footnote{In the present scenario this includes the multiplicity of potential zero eigenvalues because we assume equal-dimensional phase spaces for fermions and bosons}. This can be easily seen from the fact that the spectrum of $AB$ is identical to that of $BA$ for any two matrices.
It is also clear that the necessary and sufficient condition for a nonzero band gap (at zero energy) is $\det q(\boldsymbol{k})\neq 0$.

\section{Topological obstruction from locality and symmetry}
To discuss band topology, we should impose the locality such that $h_{\rm f,b}(\boldsymbol{k})$ is a smooth map from $T^d\equiv(2\pi \mathbb{R}/\mathbb{Z})^d$ ($d$D torus) to a matrix space constrained by the gap condition and symmetries. We define two Hamiltonians to be topologically equivalent if they can be smoothly interpolated within the map space, possibly with the assistance of some auxiliary bands \cite{Ryu2016}.

As pointed out in Ref.~\cite{Xu2020}, $h_{\rm b}(\boldsymbol{k})$ is always trivial when requiring a gap at zero energy in the dynamical matrix. This can be seen from the fact that it can always be smoothly deformed into the identity via a linear interpolation, which preserves any additional symmetries. This result is consistent with the triviality of short-range correlated bosonic Gaussian states \cite{Gong2021}. Moreover, $h_{\rm b}(\boldsymbol{k})$ can always be generated by a local and symmetric supercharge with $q(\boldsymbol{k})=\sqrt{h_{\rm b}(\boldsymbol{k})}$. 

On the other hand, $h_{\rm f}(\boldsymbol{k})$ can be nontrivial. Remarkably, if we assume that the supercharge is local and has a symmetry, we will find that not all topological fermion bands can be generated. In fact, the SUSY construction can only cover ``disentanglable'' topological phases, which can be  transformed into trivial product states \footnote{Precisely speaking, the product should be $\mathbb{Z}_2$-graded \cite{Fidkowski2011} so as to be compatible with the fermion anti-commuting relations.} by some unitary Gaussian operations that preserve locality and symmetries. 
To see this, we perform the polar decomposition of $q(\boldsymbol{k})$ in Eq.~(\ref{sck}): $q(\boldsymbol{k}) = u(\boldsymbol{k}) |q(\boldsymbol{k})|$, where  $|q(\boldsymbol{k})|\equiv\sqrt{q(\boldsymbol{k})^\dag q(\boldsymbol{k})}$ is Hermitian and positive-definite and $u(\boldsymbol{k})=q(\boldsymbol{k}) |q(\boldsymbol{k})|^{-1}$ is unitary. Without breaking locality, symmetries, or closing the gap, $q(\boldsymbol{k})$ can be smoothly unitarized into $u(\boldsymbol{k})$ via $(1-\lambda)q(\boldsymbol{k})+\lambda u(\boldsymbol{k})$ ($\lambda\in[0,1]$). Note that $u(\boldsymbol{k})Zu(\boldsymbol{k})^\dag$ generated by the unitarized supercharge is a disentanglable Hamiltonian \cite{Gong2021}. 

A broader class of topological phases can be created, if we loosen the constraints on the supercharge. If we do not require the supercharge to obey the same symmetries as the phase but still impose the locality, then those phases that become disentanglable upon removing the symmetries can be generated. In particular, all the symmetry-protected topological phases that become trivial in the absence of symmetries can arise from a local supercharge, since trivial states are in the same phase as product states and they are connected by some trivial Gaussian operations \cite{Gong2021}. We summarize the results for all the ten-fold fundamental symmetry classes in Table~\ref{table1}. While the full derivation is a bit involved (see Appendices~\ref{CSFH} and \ref{FDT1}), simple explanations are available in specific cases. For example, 2D Chern insulators (in class A) cannot be generated by local supercharges since the Wannier functions cannot be exponentially localized \cite{Marzari2007}. 

Forgoing locality as well, one can obtain all topological phases from the SUSY construction. Given an arbitrary $h_{\rm f}(\boldsymbol{k})$, the corresponding supercharge can be chosen as $q(\boldsymbol{k}) = V(\boldsymbol{k}) [\sigma_0\otimes\Lambda_+(\boldsymbol{k})^{\frac{1}{2}}]$, where $V(\boldsymbol{k})=XV(-\boldsymbol{k})^*X$ and diagonal matrix $\Lambda_+(\boldsymbol{k})>0$ are determined from $h_{\rm f}(\boldsymbol{k}) = V(\boldsymbol{k})[\sigma_z\otimes\Lambda_+(\boldsymbol{k})]V(\boldsymbol{k})^\dag$. Note that while $\Lambda_+(\boldsymbol{k})$ is always continuous in $\boldsymbol{k}$, $V(\boldsymbol{k})$ may not be continuous.

\begin{table}[tbp]
\caption{Periodic table of topological insulators and superconductors \cite{Ryu2008,Ryu2010,Kitaev2009} and their realizability by the SUSY construction. Cell uncolored: the supercharge can be local and symmetric. Cells in dark red: the supercharge necessarily breaks the locality. Cells in light red: the supercharge, if local, necessarily breaks the symmetry. Cells in light blue: a subgroup $2\mathbb{Z}$ out of $\mathbb{Z}$ can be generated by local and symmetric supercharges. Otherwise, the supercharge necessarily breaks the symmetry, provided that it is local. 
} 
\begin{center}
\begin{tabular}{ccccccccc}
\hline\hline
AZ & $d=0$ & \;\;\;\;1\;\;\;\; & \;\;\;\;2\;\;\;\; & \;\;\;\;3\;\;\;\; & \;\;\;\;4\;\;\;\; & \;\;\;\;5\;\;\;\; & \;\;\;\;6\;\;\;\; & \;\;\;\;7\;\;\;\; \\
\hline
A & \colorbox{red!10!white}{$\mathbb{Z}$} & 0 & \colorbox{red!25!white}{$\mathbb{Z}$} & 0 & \colorbox{red!10!white}{$\mathbb{Z}$} & 0 & \colorbox{red!25!white}{$\mathbb{Z}$} & 0 \\
AIII & 0 & $\mathbb{Z}$ & 0 & $\mathbb{Z}$ & 0 & $\mathbb{Z}$ & 0 & $\mathbb{Z}$ \\
\hline
AI &  \colorbox{red!10!white}{$\mathbb{Z}$}  & 0 & 0 & 0 & \colorbox{red!10!white}{$2\mathbb{Z}$} & 0 & \colorbox{red!10!white}{$\mathbb{Z}_2$} & \colorbox{red!10!white}{$\mathbb{Z}_2$} \\
BDI & $\mathbb{Z}_2$ & $\mathbb{Z}$ & 0 & 0 & 0 & $2\mathbb{Z}$ & 0 & $\mathbb{Z}_2$ \\
D & $\mathbb{Z}_2$  & $\mathbb{Z}_2$ & \colorbox{red!25!white}{$\mathbb{Z}$} & 0 & 0 & 0 & \colorbox{red!25!white}{$2\mathbb{Z}$} & 0 \\
DIII & 0 & $\mathbb{Z}_2$ & \colorbox{red!10!white}{$\mathbb{Z}_2$} & \colorbox{blue!10!white}{$\mathbb{Z}$} & 0 & 0 & 0 & $2\mathbb{Z}$ \\
AII & \colorbox{red!10!white}{$2\mathbb{Z}$} & 0 & \colorbox{red!10!white}{$\mathbb{Z}_2$} & \colorbox{red!10!white}{$\mathbb{Z}_2$} & \colorbox{red!10!white}{$\mathbb{Z}$} & 0 & 0 & 0 \\
CII & 0 & $2\mathbb{Z}$ & 0 & $\mathbb{Z}_2$ & $\mathbb{Z}_2$ & $\mathbb{Z}$ & 0 & 0 \\
C & 0 & 0 & \colorbox{red!25!white}{$2\mathbb{Z}$} & $0$ & $\mathbb{Z}_2$ & $\mathbb{Z}_2$ &  \colorbox{red!25!white}{$\mathbb{Z}$} & 0 \\
CI & 0 & 0 & 0 & $2\mathbb{Z}$ & 0 & $\mathbb{Z}_2$ & \colorbox{red!10!white}{$\mathbb{Z}_2$} & \colorbox{blue!10!white}{$\mathbb{Z}$} \\
\hline\hline
\end{tabular}
\end{center}
\label{table1}
\end{table}

\section{Identification map and entanglement duality} 
The supercharge induces a pair of identification maps which, in a canonical way, identify fermionic subsystems with bosonic subsystems, and vice versa. Subsystems identified in this way were recently shown to obey an entanglement duality \cite{Jonsson2021} which, in the following, we review briefly and investigate in the context of translation invariant systems.

The identification maps are most easily defined in terms of their action on the eigenmodes of the system. Let $\bs{\hat{f}}_{\bs{k}}=[\hat f_{\bs{k}1},\hat f_{\bs{k}2},...,\hat f_{\bs{k}|I|}]$  and $\bs{\hat{\phi}}_{\bs{k}}=[\hat\phi_{\bs{k}1},\hat\phi_{\bs{k}2},...,\hat\phi_{\bs{k}|I|}]$ be such that they diagonalize the Hamiltonians~\eqref{HfHb}, i.e., $\hat H_{\rm f} = \frac{1}{2}\sum_{\boldsymbol{k}} \sum_{\alpha=1}^{|I|} \epsilon_{ \bs{k}\alpha} (\hat{f}^\dag_{\bs{k}\alpha}  \hat{f}_{\bs{k}\alpha} + \hat{f}_{-\bs{k}\alpha} \hat{f}^\dag_{-\bs{k}\alpha})$ and $\hat H_{\rm b} = \frac{1}{2}\sum_{\boldsymbol{k}} \sum_{\alpha=1}^{|I|} \epsilon_{\bs{k}\alpha} (\hat{\phi}_{\bs k\alpha}^\dagger  \hat{\phi}_{\bs k\alpha} + \hat{\phi}_{-\bs{k}\alpha} \hat{\phi}_{-\bs{k}\alpha}^\dagger)$. Then the identification maps identify bosonic and fermionic eigenmodes with identical excitation energies. Specifically, the first map acts as $\mathcal{L}_1(\bs{\hat{f}}_{\bs{k}})=\bs{\hat{\phi}}_{\bs{k}}$ and the second acts as $\mathcal{L}_2(\bs{\hat{\phi}}_{\bs{k}})=-\ii \bs{\hat{f}}_{\bs{k}}$. In the original basis, the action of the identification map is given by
\begin{equation} 
\begin{split}
    \mathcal{L}_1\left( \begin{bmatrix}\hat{\bs c}_{\bs k}\\ \hat{\bs c}_{-\bs k}^\dagger\end{bmatrix} \right)&= 
        L_1(\bs k) \begin{bmatrix}\hat{\bs b}_{\bs k}\\ \hat{\bs b}_{-\bs k}^\dagger\end{bmatrix},\;\;\;\;
   \\
    \mathcal{L}_2\left(\begin{bmatrix}\hat{\bs b}_{\bs k}\\ \hat{\bs b}_{-\bs k}^\dagger\end{bmatrix}\right)&= 
        L_2(\bs k) \begin{bmatrix}\hat{\bs c}_{\bs k}\\ \hat{\bs c}_{-\bs k}^\dagger\end{bmatrix},
\end{split}
\end{equation}
with the matrices
\begin{equation}
    L_1\left(\bs k\right)=  |h_{\rm f}(\boldsymbol{k})|^{-\frac{1}{2}} q(\boldsymbol{k}),\;\;
    L_2(\boldsymbol{k})=-\ii  Z q(\boldsymbol{k})^\dag |h_{\rm f}(\boldsymbol{k})|^{-\frac{1}{2}}.
\end{equation}
Hence, if the fermion-boson coupling is short-ranged, \ie $q(\bs k)$ is analytic, so are $L_1(\bs k)$ and $L_2(\bs k)$. Therefore, for a short-range supercharge $\hat Q$ both $\mathcal{L}_1$ and $\mathcal{L}_2$ are locality preserving mappings between the fermionic and  bosonic lattices.

The identification maps preserve the expectation value of the commutator and anti-commutator of linear operators, i.e., for arbitrary fermionic linear operators $\hat f_1, \hat f_2$
\begin{align}\label{eq:comm_preserve}
    \bra{\Psi_\ff}[\hat f_1,\hat f_2]\ket{\Psi_\ff} = \bra{\Psi_\bb}[\mathcal L_1(\hat f_1),\mathcal L_1(\hat f_2)]\ket{\Psi_\bb} ,
\end{align}
and accordingly for the anti-commutator, and for $\mathcal L_2$. Also, the identification maps fully encode the ground state of the Hamiltonians~\eqref{sck}. 
This is because their compositions yield the linear complex structures 
    $\mathcal{J}_\ff=\mathcal{L}_2\circ \mathcal{L}_1$ of the fermionic ground state $\ket{\Psi_\ff}$, and
    $\mathcal{J}_\bb= \mathcal{L}_1\circ \mathcal{L}_2$ of the bosonic ground state $\ket{\Psi_\bb}$ \cite{hackl_bosonic_2021,hackl_minimal_2019,Jonsson2021}. The linear complex structures relate the expectation values of anti-commutators and commutators in the ground states as $\bra{\Psi_\bb}\{\hat \phi_1,\hat \phi_2\}\ket{\Psi_\bb} = [\hat \phi_1,\mathcal{J}_\bb(\hat \phi_2)]$ for any bosonic linear operators $\hat \phi_1,\hat \phi_2$,
and $\bra{\Psi_\ff}[\hat f_1,\hat f_2]\ket{\Psi_\ff} =-\{\hat f_1, \mathcal{J}_\ff(\hat f_2)\}$ for any fermionic linear operators $\hat f_1,\hat f_2$.

The linear complex structures also encode the entanglement content of the ground states. 
To this end, consider subsystems consisting of a single fermionic~/~bosonic mode $\hat c$~/~$\hat b$. If the reduced state of a mode is pure, \ie if the mode is in a product state with the complement of its ground state $\ket{\Psi_{\ff/\bb}}$, then the restriction of $\mathcal{J}_{\ff/\bb}$ to the linear space spanned by $\hat c$~/~$\hat b$ and $\hat c^\dagger$~/~$\hat b^\dag$ has eigenvalues $\pm \ii$ \footnote{The eigenvalues are associated with the restricted representation matrix of $\mathcal{J}_{\rm f/b}$.}.
However, if the reduced state is mixed, \ie the mode is entangled with the complement, then the restriction of $\mathcal{J}_{\ff/\bb}$ to the subsystem has eigenvalues $\pm\ii\lambda_{\ff/\bb}$ with $\lambda_\ff\in[0,1)$, $\lambda_\bb\in(1,\infty]$. The von Neumann entropy of the mode is then $s(\lambda_{\ff/\bb})$ with $s(x)=\frac{1+x}{2}|\log(\frac{1+x}{2})|-\frac{|1-x|}{2}\log(\frac{|1-x|}{2})$. For fermions, $\lambda_\ff\to0$ signals maximal entanglement of $s_\ff\to\log2$, and for bosons, $\lambda_\ff\to0$ signals diverging entanglement $s_\bb\to\infty$. Since $|\Psi_{\rm f/b}\rangle$ is Gaussian, subsystems consisting of several modes can be decomposed in terms of the subsystems' normal modes and then  the above applies mode by mode \cite{hackl_bosonic_2021,hackl_minimal_2019,Jonsson2021}.

The entanglement duality \cite{Jonsson2021} asserts that if a fermionic and a bosonic  subsystem are related by an identification map, \eg $\mathcal{L}_1(\hat c)=\hat b$ or $\mathcal{L}_2(\hat b)=\hat c$, then the eigenvalues of the restricted complex linear structures are inverses of each other,
\begin{equation}\label{eq:inverse_relation_duality}
    \lambda_\ff=1/{\lambda_\bb}.
\end{equation}
This means that, on the one hand, modes in pure states are mapped to modes in pure states, and, on the other hand, highly entangled modes are mapped to highly entangled modes. 

If the fermion system is topological, the reduced state on some region of the lattice is known to contain some (almost) maximally entangled modes~\cite{Fidkowski2010}. For such modes the eigenvalue $\lambda_\ff$ approaches zero and its entanglement entropy approaches $\log 2$.
On the bosonic side, such a system is mapped to a subsystem with divergent values of $\lambda_\bb$, and thus a divergent entanglement entropy.
However, because the identification maps preserve locality, the real-space profile does not change much under the mapping from the fermionic to the bosonic lattice. Hence, the divergent entanglement in the dual bosonic mode necessarily is due to high squeezing.
In fact, this can be seen by considering a highly entangled fermionic mode operator $\hat c'$, \ie with 
$\bra{\Psi_\ff}[\hat c',\hat c'^\dagger]\ket{\Psi_\ff} =\epsilon\ll 1$.
Through $\mathcal L_1$ this mode is identified with the bosonic mode defined by $\hat b'=\mathcal L\left(\hat c'\right)/\sqrt\epsilon$. The rescaling with $\sqrt\epsilon$, which is necessary due to~\eqref{eq:comm_preserve} to obtain a proper bosonic normalization, causes the coefficients in the expansion of $\hat b'$  to diverge  as $\epsilon\to0$.

\begin{figure}
	\includegraphics[width=8.4cm, clip]{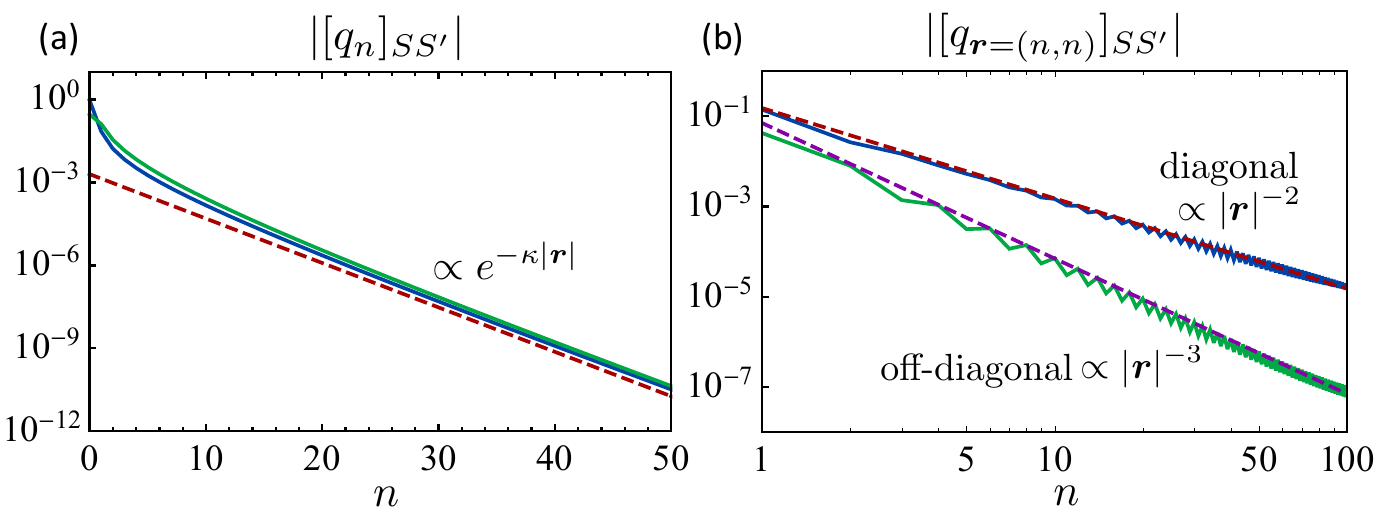}
	\caption{Real-space profiles of the diagonal (blue) and off-diagonal (green) components in the supercharge for (a) the Kitaev chain (\ref{KC}) with $\mu=1$, $t=0.7$ (system size: $400$) and (b) the 2D chiral superconductor (\ref{cSC}) with $m=1$ (direction: $\boldsymbol{r}\propto(1,1)$, system size: $400\times 400$). In (a), both components decay exponentially. In (b), the diagonal/off-diagonal decays algebraically as $|\boldsymbol{r}|^{-2}$ / $|\boldsymbol{r}|^{-3}$. Here $q_{\boldsymbol{r}}= |\Omega|^{-1}\sum_{\boldsymbol{k}} q(\boldsymbol{k})e^{i\boldsymbol{k}\cdot\boldsymbol{r}}$ and $S=\pm s$ ($s\in I$), where ``$\pm$" arises from the creation or annihilation half of the basis [cf. Eq.~(\ref{sck})].}
	\label{fig:decay}
\end{figure}

\section{Examples}
As a canonical example we consider the 1D Kitaev chain \cite{Kitaev2001} under periodic boundary conditions:
\begin{align}
h_{\rm f}(k)= -2t\sin k \sigma_y + (\mu+2t\cos k)\sigma_z.
\label{KC}
\end{align}
Following the construction given in the Appendices \ref{CSFH} and \ref{PCGT}, one finds that this Hamiltonian is generated by the supercharge $q(k)=\sqrt{\epsilon_k}V(k)$ with $\epsilon_k=\sqrt{\mu^2+4t^2+4t\mu\cos k}$ and $V(k)=(\sigma_0+\sigma_x)/2 + (\mu+2te^{\ii k})(\sigma_0-\sigma_x)/(2\epsilon_k)$. This supercharge generates the bosonic Hamiltonian with $h_\bb(k)=\epsilon_k\sigma_0$, which, in contrast to the Kitaev chain, is particle-number conserving, and thus has a trivial ground state.

\begin{figure}
\begin{center}
  \includegraphics[width=\linewidth]{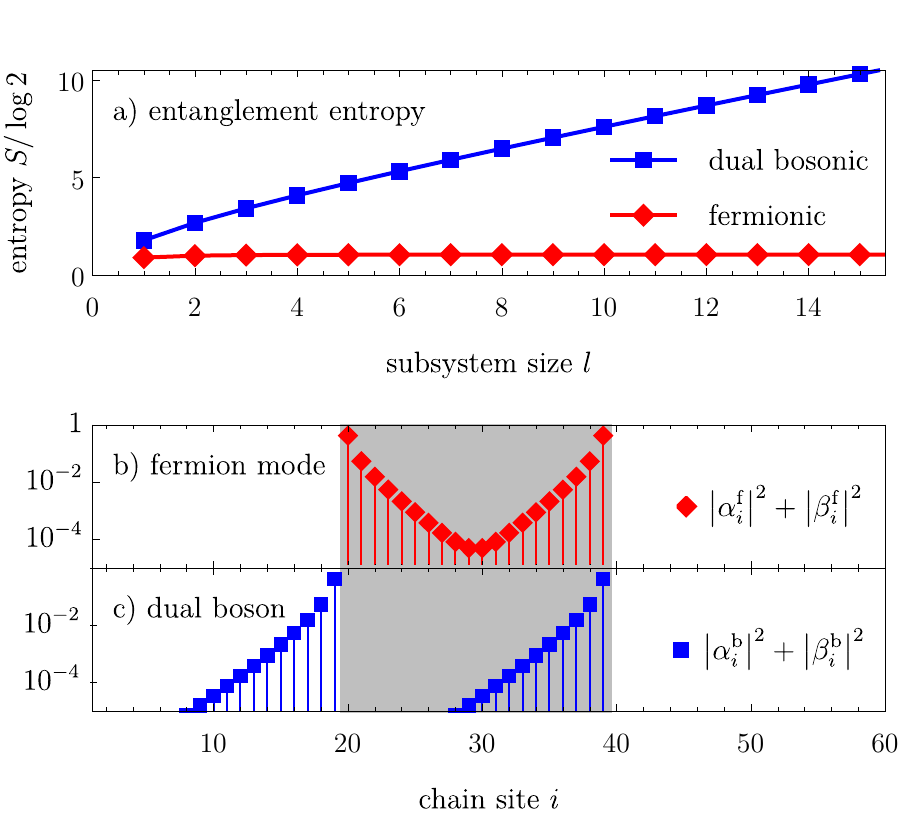}
\end{center}\vspace{-5mm}
\caption{Entanglement in the fermionic 1D Kitaev chain~\eqref{KC}  with $\mu=1$ and $t=0.7$ (system size: $60$) and its SUSY bosonic dual. (a) Entanglement entropy for fermionic subsystems of varying length $l$, and for their dual bosonic subsystems. (b) Profile of a topological fermionic entanglement edge mode 
$\hat{c}'=\sum_i\alpha^\ff_i \hat c_i+\beta^\ff_i\hat c^\dagger_i$ carrying almost maximal entanglement. This mode is symmetrically localized at both edges of the subsystem (shaded region). (c) Profile of the dual bosonic mode $\mathcal{L}_1\left(\hat c'\right)=\sum_i\alpha^\bb_i \hat c_i+\beta^\bb_i\hat c^\dagger_i$. In contrast to the fermionic mode, it is not mirror-symmetric and, counter-intuitively, at the left edge it is localized outside the shaded region. (Note that $\mathcal{L}_1\left(\hat c'\right)$ is not a normalized bosonic ladder operator, but $[\mathcal{L}_1\left(\hat c'\right),\mathcal{L}_1\left(\hat c'\right)^\dagger]\approx 1.78\cdot 10^{-4}$, hence after normalizing the operator we obtain a highly squeezed bosonic mode.)
} 
\label{fig:kitaevchain} 
\end{figure}

Both the supercharge [cf. Fig.~\ref{fig:decay}(a)] and the bosonic Hamiltonian are short range. However, whereas both $\hat H_{\rm f}$ and $\hat H_{\rm b}$ are mirror-symmetric, the supercharge is not. This asymmetry shows most clearly in the flat band case, $\mu=0$, where the fermionic eigenmodes are formed by pairing Majorana modes on adjacent sides, $\hat{f}_j = (\hat c_j+\hat c_{j+1}+\hat c_j^\dagger - \hat c_{j+1}^\dagger)/2$.
Note that $\mathcal{L}_1(\hat f_j)=\hat b_j$, implying that the bosonic operator on the $j$th site is identified with a fermionic mode residing equally on the $j$th and $(j+1)$th sites, thus breaking mirror-symmetry. 

This asymmetry persists in the topological phase of the Kitaev chain also for non-zero values of $\mu$. Fig.~\ref{fig:kitaevchain} demonstrates this by showing the localization of the fermionic entanglement edge mode in a subsystem and that of the bosonic dual mode. The figure also conveys that the dual bosonic mode is highly squeezed. This behaviour is also related to the very different scaling of the entanglement entropy of fermionic subsytems and their bosonic dual subsystems, also given in Fig.~\ref{fig:kitaevchain}. On the fermion side, the entanglement entropy approaches a finite value as  the subsystem size  $l$ increases. Meanwhile, as $l$ increases, the minimal $\lambda_\ff$ decays exponentially. Due to the duality~\eqref{eq:inverse_relation_duality}, this leads to dual bosonic eigenvalue which is increasing exponentially, thus leading to a scaling of $\mathcal O(l)$ of the entanglement entropy on the boson side. 

We turn to consider a 2D chiral superconductor (class D), which is also described by a two-band BdG Hamiltonian 
\begin{equation}
h_{\rm f}(\boldsymbol{k})=\sin k_x\sigma_x + \sin k_y\sigma_y + (m - \cos k_x - \cos k_y)\sigma_z. 
\label{cSC}
\end{equation}
For $0< |m|<2$, the system is in a topological phase characterized by a nontrivial Chern number \cite{Bernevig2013}. According to the previous general analysis, such a topological phase cannot be generated by a local supercharge. Nevertheless, we can still construct a nonlocal supercharge whose diagonal (hopping)/off-diagonal (pairing) component follows a power-law decay $|\boldsymbol{r}|^{-2}$/$|\boldsymbol{r}|^{-3}$ [cf. Fig.~\ref{fig:decay}b]. These algebraic tails arise from the non-analyticity of $q(\boldsymbol{k})$ at specific points in the Brillouin zone (see Appendix~\ref{GC}). It might be interesting to ask whether the supercharge could be more localized than $|\boldsymbol{r}|^{-2}$ decay. 

As a final example, we mention that all the time-reversal-symmetric topological insulators (class AII), both in 2D and 3D, can be generated by local time-reversal-symmetric supercharges that break the ${\rm U}(1)$ symmetry (see Appendix~\ref{AIAII}).

\section{Discussions}
\emph{Strictly} local systems, whose coupling ranges are finite, constitute an important subclass of short-range systems. In particular, all the examples  of $\hat H_{\rm f}$ above fall into this category. It is natural to ask whether the supercharge can also be chosen to be strictly local for them. While we do not have a complete answer, we can show this is possible at least for classes BDI, CII and AIII in any dimensions (see Appendix~\ref{chiral}). This is to be contrasted to zero-length correlated topological phases, i.e., those with  strictly local (compactly supported) Wannier functions, which do not exist in $d\ge2$D for all the fundamental symmetry classes \cite{Read2017}. 

A remarkable observation in the numerical demonstration is the spatial asymmetry in the identification map for the Kitaev chain, which is mirror-symmetric. In fact, the identification map, or equivalently the supercharge can never be mirror-symmetric under the locality constraint. The most convenient way to see this is to forget the time-reversal symmetry and regard the Kitaev chain as a nontrivial phase in class D, which is classified by $\mathbb{Z}_2$ in 1D. This $\mathbb{Z}_2$ index is determined by the parity of the winding number of $q(k)$. Provided that the supercharge is mirror-symmetric, its winding number will be enforced to be even and thus cannot generate a nontrivial Kitaev chain. It would be interesting to explore the topological constraints of additional symmetries in a more general setting.

Finally, we note that the exemplified boson Hamiltonians respect ${\rm U}(1)$ particle-number symmetry, although the corresponding fermion Hamiltonians and supercharges do not. In fact, given a general local supercharge, we can always gauge transform it in a locality-preserving manner such that $h_{\rm f}(\boldsymbol{k})$ ($h_{\rm b}(\boldsymbol{k})$) remains invariant, while $h_{\rm b}(\boldsymbol{k})$ ($h_{\rm f}(\boldsymbol{k})$) becomes particle-conserving (see Appendix~\ref{PCGT}). In particular, this result implies that all the fermion topological phases can be mapped into the boson vacuum, the ground state of an arbitrary ${\rm U}(1)$ symmetric boson Hamiltonian.

\section{Summary and outlook}
We have examined the role of topology in the SUSY construction of quadratic fermion and boson Hamiltonians. We have found that not all the topological fermion bands can be generated by local and symmetric supercharges. We have also clarified that the boson bands are always trivial and the topological information is encoded in the SUSY map. The apparent inconsistency with the entanglement duality can be resolved by noting that the bosonic subsystem could be highly squeezed.

Our work raises many open problems for future studies. On top of those in the discussion part, it would be interesting to see whether and how the SUSY construction can be extended to unstable and meta-stable \cite{Clerk2018,Viola2021} bosonic systems. Probably the supercharge and the fermion Hamiltonian should be non-Hermitian \cite{Ashida2021,Kunst2011}. 
It might also be interesting to consider the generalization to a full open-system setting (described by Lindbladians) and to systems in the continuum rather than lattices.

\acknowledgements
We thank Ignacio Cirac and Krishanu Roychowdhury for helpful discussions. Z.G. is supported by the Max-Planck-Harvard Research Center for Quantum Optics (MPHQ). R.H.J. gratefully acknowledges support by the Wenner-Gren Foundations. D.M. acknowledges funding from ERC Advanced Grant QUENOCOBA under the EU Horizon 2020 program (Grant Agreement No. 742102).

\appendix


\section{Derivation of the SUSY Hamiltonians}
\label{DSUSYH}
\subsection{Supercharge}
We start from explaining why a translation-invariant supercharge takes the form of Eq.~(\ref{sck}). In the most general case, a supercharge for $N$ pair of fermion and boson modes, denoted as $c_j$ and $b_j$ ($j=1,2,...,N$), respectively, can be written as
\begin{equation}
\hat Q = \begin{bmatrix} \hat{\boldsymbol{c}}^\dag \\ \hat{\boldsymbol{c}} \end{bmatrix}^{\rm T} \begin{bmatrix} U^* & T^* \\ T & U \end{bmatrix}  \begin{bmatrix} \hat{\boldsymbol{b}} \\ \hat{\boldsymbol{b}}^\dag \end{bmatrix}
\label{SC}
\end{equation}
where $\hat{\boldsymbol{c}} = [\hat c_1,\hat c_2,...,\hat c_N]^{\rm T}$ and $\hat{\boldsymbol{b}}=[\hat b_1,\hat b_2,...,\hat b_N]^{\rm T}$ and $U$, $T$ are two arbitrary $N\times N$ matrices. 

Let us then consider the specific situation in the main text, i.e., both of the fermions and bosons live on a $d$D lattice $\Omega\subset \mathbb{Z}^d$ with a set of internal states $I$ at each unit cell. In this case, we have $N=|\Omega||I|$ and the fermion/boson modes are denoted as $c_{\boldsymbol{r}s}$/$b_{\boldsymbol{r}s}$ ($\boldsymbol{r}\in\Omega$, $s\in I$). Imposing periodic boundary conditions and translation invariance on the supercharge (\ref{SC}), we have $T_{\boldsymbol{r}s,\boldsymbol{r}'s'} = T_{\boldsymbol{r}-\boldsymbol{r}',ss'}$ and $U_{\boldsymbol{r}s,\boldsymbol{r}'s'} = U_{\boldsymbol{r}-\boldsymbol{r}',ss'}$ and can define their Fourier transformations as $[T_{\boldsymbol{k}}]_{ss'}= \sum_{\delta\boldsymbol{r}\in\Omega}T_{\delta\boldsymbol{r},ss'}e^{-\ii\boldsymbol{k}\cdot\delta\boldsymbol{r}}$, $[U_{\boldsymbol{k}}]_{ss'}= \sum_{\delta\boldsymbol{r}\in\Omega}U_{\delta\boldsymbol{r},ss'}e^{-\ii\boldsymbol{k}\cdot\delta\boldsymbol{r}}$, which are both $|I|\times|I|$ matrices. In terms of the momentum basis, we can decompose Eq.~(\ref{SC}) as $\hat Q=\sum_{\boldsymbol{k}} \hat Q_{\boldsymbol{k}}$, where
\begin{equation}
\hat Q_{\boldsymbol{k}} = \begin{bmatrix} \hat{\boldsymbol{c}}^\dag_{\boldsymbol{k}} \\ \hat{\boldsymbol{c}}_{-\boldsymbol{k}} \end{bmatrix}^{\rm T} \begin{bmatrix} U^*_{-\boldsymbol{k}} & T^*_{-\boldsymbol{k}} \\ T_{\boldsymbol{k}} & U_{\boldsymbol{k}} \end{bmatrix}  \begin{bmatrix} \hat{\boldsymbol{b}}_{\boldsymbol{k}} \\ \hat{\boldsymbol{b}}^\dag_{-\boldsymbol{k}} \end{bmatrix}.
\label{Qk}
\end{equation}
The $2|I|\times 2|I|$ matrix in Eq.~(\ref{Qk}) turns out to be the general form of $q(\boldsymbol{k})$ constrained by Eq.~(\ref{qPHS}). One can also check that 
\begin{equation}
\hat Q_{\boldsymbol{k}}^\dag = \hat Q_{-\boldsymbol{k}},\;\;\;\;\{\hat Q_{\boldsymbol{k}},\hat Q_{\boldsymbol{k}'}\}=0,\;\;\forall\boldsymbol{k}\neq-\boldsymbol{k}'.
\label{QkQk}
\end{equation}

\subsection{BdG Hamiltonians}
To simplify the calculations, we can express the SUSY Hamiltonian as an anti-commutator:
\begin{equation}
\hat H = \hat Q^2 = \frac{1}{2}\{\hat Q,\hat Q\}.
\end{equation}
Using Eq.~(\ref{QkQk}), we can decompose the SUSY Hamiltonian as $\hat H=\sum_{\boldsymbol{k}} \hat H_{\boldsymbol{k}}$, where
\begin{equation}
\hat H_{\boldsymbol{k}}=\frac{1}{2}\{\hat Q_{\boldsymbol{k}},\hat Q_{-\boldsymbol{k}}\}=\frac{1}{2}\{\hat Q_{\boldsymbol{k}},\hat Q^\dag_{\boldsymbol{k}}\}.
\end{equation}
Noting that
\begin{equation}
\hat Q_{\boldsymbol{k}} = \hat r_{\boldsymbol{k}} + \hat r^\dag_{-\boldsymbol{k}},\;\;\;\;\hat r_{\boldsymbol{k}} = 
\hat{\boldsymbol{c}}^\dag_{\boldsymbol{k}} (U^*_{-\boldsymbol{k}} \hat{\boldsymbol{b}}_{\boldsymbol{k}} + T^*_{-\boldsymbol{k}} \hat{\boldsymbol{b}}^\dag_{-\boldsymbol{k}}),
\end{equation}
we can express $H_{\boldsymbol{k}}$ explicitly as
\begin{equation}
\hat H_{\boldsymbol{k}}= \frac{1}{2}[\{\hat r_{\boldsymbol{k}},\hat r_{-\boldsymbol{k}}\} + \{\hat r_{\boldsymbol{k}},\hat r^\dag_{\boldsymbol{k}}\} + \{\hat r^\dag_{\boldsymbol{k}},\hat r^\dag_{-\boldsymbol{k}}\} + \{\hat r_{-\boldsymbol{k}},\hat r^\dag_{-\boldsymbol{k}}\}].
\end{equation}
Hence, it suffices to calculate $\{\hat r_{\boldsymbol{k}},\hat r_{-\boldsymbol{k}}\}$ and $\{\hat r_{\boldsymbol{k}},\hat r^\dag_{\boldsymbol{k}}\}$, since we can then obtain $\{\hat r_{\boldsymbol{k}}^\dag,\hat r^\dag_{-\boldsymbol{k}}\}$ and $\{\hat r_{-\boldsymbol{k}},\hat r^\dag_{-\boldsymbol{k}}\}$ by taking the Hermitian conjugate or replacing $\boldsymbol{k}$ by $-\boldsymbol{k}$, respectively.

The result of $\{\hat r_{\boldsymbol{k}},\hat r_{-\boldsymbol{k}}\}$ turns out to be
\begin{equation}
\{\hat r_{\boldsymbol{k}},\hat r_{-\boldsymbol{k}}\} = [\hat{\boldsymbol{c}}^\dag_{\boldsymbol{k}}]^{\rm T}(U^*_{-\boldsymbol{k}} T^\dag_{\boldsymbol{k}} - T^*_{-\boldsymbol{k}} U^\dag_{\boldsymbol{k}}) \hat{\boldsymbol{c}}^\dag_{-\boldsymbol{k}},
\end{equation}
from which we know
\begin{equation}
\{\hat r^\dag_{\boldsymbol{k}},\hat r^\dag_{-\boldsymbol{k}}\} = \hat{\boldsymbol{c}}_{-\boldsymbol{k}}^{\rm T}(T_{\boldsymbol{k}}U^{\rm T}_{-\boldsymbol{k}} - U_{\boldsymbol{k}}T^{\rm T}_{-\boldsymbol{k}}) \hat{\boldsymbol{c}}_{\boldsymbol{k}}.
\end{equation}
The result of $\{\hat r_{\boldsymbol{k}},\hat r^\dag_{\boldsymbol{k}}\}$ turns out to be
\begin{equation}
\begin{split}
\{\hat r_{\boldsymbol{k}},\hat r^\dag_{\boldsymbol{k}}\} & = [\hat{\boldsymbol{b}}^\dag_{\boldsymbol{k}}]^{\rm T}U^{\rm T}_{-\boldsymbol{k}}U^*_{-\boldsymbol{k}}\hat{\boldsymbol{b}}_{\boldsymbol{k}} + \hat{\boldsymbol{b}}_{-\boldsymbol{k}}^{\rm T}T^{\rm T}_{-\boldsymbol{k}}T^*_{-\boldsymbol{k}}\hat{\boldsymbol{b}}^\dag_{-\boldsymbol{k}} \\
&+ \hat{\boldsymbol{b}}_{-\boldsymbol{k}}^{\rm T} T^{\rm T}_{-\boldsymbol{k}} U^*_{-\boldsymbol{k}}\hat{\boldsymbol{b}}_{\boldsymbol{k}} + [\hat{\boldsymbol{b}}^\dag_{\boldsymbol{k}}]^{\rm T} U^{\rm T}_{-\boldsymbol{k}} T^*_{-\boldsymbol{k}}\hat{\boldsymbol{b}}^\dag_{-\boldsymbol{k}} \\
&+ [\hat{\boldsymbol{c}}^\dag_{\boldsymbol{k}}]^{\rm T}(U^*_{-\boldsymbol{k}}U^{\rm T}_{-\boldsymbol{k}} - T^*_{-\boldsymbol{k}}T^{\rm T}_{-\boldsymbol{k}})\hat{\boldsymbol{c}}_{\boldsymbol{k}},
\end{split}
\end{equation}
from which we know
\begin{equation}
\begin{split}
\{\hat r_{-\boldsymbol{k}},\hat r^\dag_{-\boldsymbol{k}}\} & = \hat{\boldsymbol{b}}_{-\boldsymbol{k}}^{\rm T}U^\dag_{\boldsymbol{k}}U_{\boldsymbol{k}}\hat{\boldsymbol{b}}^\dag_{-\boldsymbol{k}} + [\hat{\boldsymbol{b}}^\dag_{\boldsymbol{k}}]^{\rm T}T^\dag_{\boldsymbol{k}}T_{\boldsymbol{k}}\hat{\boldsymbol{b}}_{\boldsymbol{k}} \\
&+ \hat{\boldsymbol{b}}^{\rm T}_{-\boldsymbol{k}} U^\dag_{\boldsymbol{k}}T_{\boldsymbol{k}} \hat{\boldsymbol{b}}_{\boldsymbol{k}} + [\boldsymbol{b}^\dag_{\boldsymbol{k}}]^{\rm T} T^\dag_{\boldsymbol{k}}U_{\boldsymbol{k}} \boldsymbol{b}^\dag_{-\boldsymbol{k}} \\
& + \hat{\boldsymbol{c}}_{-\boldsymbol{k}}^{\rm T}(T_{\boldsymbol{k}}T^\dag_{\boldsymbol{k}} - U_{\boldsymbol{k}}U^\dag_{\boldsymbol{k}} )\hat{\boldsymbol{c}}^\dag_{-\boldsymbol{k}}.
\end{split}
\end{equation}
Combining all the results above, we end up with
\begin{equation}
\begin{split}
\hat H^{\rm f}_{\boldsymbol{k}}&= \frac{1}{2} \begin{bmatrix} \hat{\boldsymbol{c}}^\dag_{\boldsymbol{k}} \\ \hat{\boldsymbol{c}}_{-\boldsymbol{k}} \end{bmatrix}^{\rm T} h_{\rm f}(\boldsymbol{k}) \begin{bmatrix} \hat{\boldsymbol{c}}_{\boldsymbol{k}} \\ \hat{\boldsymbol{c}}^\dag_{-\boldsymbol{k}} \end{bmatrix},\\
h_{\rm f}(\boldsymbol{k})&= 
\begin{bmatrix} 
U^*_{-\boldsymbol{k}}U^{\rm T}_{-\boldsymbol{k}} - T^*_{-\boldsymbol{k}}T^{\rm T}_{-\boldsymbol{k}} & U^*_{-\boldsymbol{k}} T^\dag_{\boldsymbol{k}} - T^*_{-\boldsymbol{k}} U^\dag_{\boldsymbol{k}} \\ 
T_{\boldsymbol{k}}U^{\rm T}_{-\boldsymbol{k}} - U_{\boldsymbol{k}}T^{\rm T}_{-\boldsymbol{k}} & T_{\boldsymbol{k}}T^\dag_{\boldsymbol{k}} - U_{\boldsymbol{k}}U^\dag_{\boldsymbol{k}}
\end{bmatrix},  
\end{split}
\label{hf}
\end{equation}
and
\begin{equation}
\begin{split}
\hat H^{\rm b}_{\boldsymbol{k}}&= \frac{1}{2}\begin{bmatrix} \hat{\boldsymbol{b}}^\dag_{\boldsymbol{k}} \\ \hat{\boldsymbol{b}}_{-\boldsymbol{k}} \end{bmatrix}^{\rm T} h_{\rm b}(\boldsymbol{k})
\begin{bmatrix} \hat{\boldsymbol{b}}_{\boldsymbol{k}} \\ \hat{\boldsymbol{b}}^\dag_{-\boldsymbol{k}} \end{bmatrix},\\
h_{\rm b}(\boldsymbol{k})&=
\begin{bmatrix} 
U^{\rm T}_{-\boldsymbol{k}}U^*_{-\boldsymbol{k}} + T^\dag_{\boldsymbol{k}}T_{\boldsymbol{k}} & U^{\rm T}_{-\boldsymbol{k}} T^*_{-\boldsymbol{k}} + T^\dag_{\boldsymbol{k}}U_{\boldsymbol{k}} \\ 
T^{\rm T}_{-\boldsymbol{k}} U^*_{-\boldsymbol{k}} + U^\dag_{\boldsymbol{k}}T_{\boldsymbol{k}} & T^{\rm T}_{-\boldsymbol{k}}T^*_{-\boldsymbol{k}} + U^\dag_{\boldsymbol{k}}U_{\boldsymbol{k}}
\end{bmatrix}.  
\end{split}
\label{hb}
\end{equation}
With the matrix in Eq.~(\ref{Qk}) denoted as $q(\boldsymbol{k})$, one can check that Eqs.~(\ref{hf}) and (\ref{hb}) are nothing but Eq.~(\ref{hfhb}).

\section{Constructing supercharges out of fermion Hamiltonians}
\label{CSFH}

\subsection{General construction}
\label{GC}
We introduce a general and simple way to extract the supercharge and construct the SUSY boson band for a given gapped fermion BdG Hamiltonian $h_{\rm f} (\boldsymbol{k})$, although this construction may not guarantee the locality. 

Having in mind that $h_{\rm f} (\boldsymbol{k})$ is gapped and respects the particle-hole symmetry (i.e., $Xh_{\rm f} (\boldsymbol{k})^* X = - h_{\rm f} (-\boldsymbol{k})$), we know that there should be exactly $n=|I|$ positive eigenenergy $\epsilon_{\boldsymbol{k}\alpha}>0$ for each $\boldsymbol{k}$. We can thus identify the corresponding normalized eigenstates:
\begin{equation}
h_{\rm f} (\boldsymbol{k}) u_{\boldsymbol{k}\alpha} = \epsilon_{\boldsymbol{k}\alpha} u_{\boldsymbol{k}\alpha},\;\;\;\;\alpha=1,2,...,n.
\end{equation}
By again using the PHS, we have
\begin{equation}
Xh_{\rm f} (-\boldsymbol{k})^* u_{-\boldsymbol{k}\alpha}^* = -h_{\rm f} (\boldsymbol{k})X u_{-\boldsymbol{k}\alpha}^* 
=  \epsilon_{-\boldsymbol{k}\alpha} X u_{-\boldsymbol{k}\alpha}^*,
\end{equation}
implying that $Xu_{-\boldsymbol{k}\alpha}^*$ is also an eigenstate but with a negative eigenenergy $-\epsilon_{-\boldsymbol{k}\alpha}$. Obviously, $\{u_{\boldsymbol{k}\alpha}, Xu^*_{-\boldsymbol{k}\alpha}\}^n_{\alpha=1}$ form the complete eigenbasis of $h_{\rm f} (\boldsymbol{k})$, indicating the following spectral decomposition (diagonalization):
\begin{equation}
\begin{split}
&h_{\rm f} (\boldsymbol{k})V(\boldsymbol{k}) = V(\boldsymbol{k}) \Lambda(\boldsymbol{k}),\\
&V(\boldsymbol{k})=\begin{bmatrix} 
\uparrow 
& \cdots & \uparrow & \uparrow & \cdots & \uparrow \\
u_{\boldsymbol{k}1} 
& \cdots & u_{\boldsymbol{k}n} & Xu_{-\boldsymbol{k}1}^* & ... & Xu_{-\boldsymbol{k}n}^* \\
\downarrow 
& \cdots & \downarrow & \downarrow & \cdots & \downarrow
\end{bmatrix}, \\
&\Lambda(\boldsymbol{k})={\rm diag}
\begin{bmatrix} \epsilon_{\boldsymbol{k}1} 
& \cdots & \epsilon_{\boldsymbol{k}n} & - \epsilon_{-\boldsymbol{k}1} 
& \cdots & - \epsilon_{-\boldsymbol{k}n} \end{bmatrix}. 
\end{split}
\end{equation}
Such a choice of the eigenbasis has appeared in Ref.~\cite{Teo2010} (cf. Eq.~(5.2)) and plays an important role in calculating PHS-protected topological invariants. The privilege of this eigenbasis is rooted in the following symmetry property:
\begin{equation}
\begin{split}
&XV(\boldsymbol{k})^* X =V(-\boldsymbol{k}),\\
&X|\Lambda(\boldsymbol{k})|X=|\Lambda(-\boldsymbol{k})|,\;\;
|\Lambda(\boldsymbol{k})|=Z\Lambda(\boldsymbol{k}). 
\end{split}
\end{equation}
Accordingly, we can choose $q(\boldsymbol{k})$ to be
\begin{equation}
q(\boldsymbol{k}) = V(\boldsymbol{k})|\Lambda(\boldsymbol{k})|^{1/2},
\label{qVL}
\end{equation}
such that Eq.~(\ref{qPHS}) is fulfilled. For a general two-band BdG Hamiltonian $h_{\rm f}(\boldsymbol{k})=\sum_{\mu=x,y,z} d_\mu(\boldsymbol{k})\sigma_\mu$, one can check that a valid choice reads
\begin{equation}
q =\frac{1}{\sqrt{2}}
\begin{pmatrix}
e^{-i\phi}\sqrt{|\boldsymbol{d}| + d_z} & \sqrt{|\boldsymbol{d}| - d_z} \\
\sqrt{|\boldsymbol{d}| - d_z} & -e^{i\phi}\sqrt{|\boldsymbol{d}| + d_z}
\end{pmatrix},
\end{equation}
where variable $\boldsymbol{k}$ is dropped for simplicity, $|\boldsymbol{d}|=\sqrt{d_x^2+d_y^2+d_z^2}$, $e^{i\phi}=(d_x + id_y)/\sqrt{d_x^2 + d_y^2}$, $d_\mu(\boldsymbol{k})$'s are real and satisfy $d_{x,y}(-\boldsymbol{k})=-d_{x,y}(\boldsymbol{k})$, $d_z(-\boldsymbol{k}) = d_z(\boldsymbol{k})$. Note that the (non-local) supercharge for the chiral superconductor in the main text follows this construction. For this example, one can check that $e^{i\phi}$ is not well-defined at the high-symmetry points $\Gamma=(0,0),(0,\pi),(\pi,0)$ and $(\pi,\pi)$, while $\sqrt{|\boldsymbol{d}|\pm d_z}$ is well-defined but may exhibit a linear singularity $\sim|\boldsymbol{k}-\Gamma|$ nearby. After the Fourier transform, these singularities are turned into the algebraic tails of the supercharge in real space, as shown in Fig.~\ref{fig:decay}(b). Similar phenomena have been observed for the parent Hamiltonians of chiral Gaussian fermionic projected entangled pair states~\cite{Wahl2014}.

Remarkably, the corresponding boson BdG Hamiltonian of Eq.~(\ref{qVL}) is diagonalized:
\begin{equation}
h_{\rm b}(\boldsymbol{k}) = q(\boldsymbol{k})^\dag q(\boldsymbol{k}) = |\Lambda(\boldsymbol{k})|.
\label{diag}
\end{equation}
This Hamiltonian is always local and conserves the particle number. One can also easily write down the identification maps:
\begin{equation}
L_1(\boldsymbol{k}) = V(\boldsymbol{k}),\;\;\;\;L_2(\boldsymbol{k}) = Z V(\boldsymbol{k})^\dag,
\end{equation}
which are both unitary.

While the above general construction may not be local, we can actually construct a local supercharge from another local one for a different Hamiltonian, provided that Hamiltonian is in the same (in the homotopy sense) topological phase as the target one. To see this, we consider $h_{\rm f}(\boldsymbol{k};0)$ and $h_{\rm f}(\boldsymbol{k};1)$ which can be smoothly interpolated by $h_{\rm f}(\boldsymbol{k};\lambda)$ ($\lambda\in[0,1]$). Suppose that $h_{\rm f}(\boldsymbol{k};0)$ can be generated by a local supercharge $q_0(\boldsymbol{k})$, then we can construct $q_1(\boldsymbol{k})=v_1(\boldsymbol{k})q_0(\boldsymbol{k})$ that generates $h_{\rm f}(\boldsymbol{k};1)$, where $v_1(\boldsymbol{k})$ is obtained by solving 
\begin{equation}
\partial_\lambda v_\lambda(\boldsymbol{k}) = \frac{1}{2}[\partial_\lambda h_{\rm f}(\boldsymbol{k};\lambda)] h_{\rm f}(\boldsymbol{k};\lambda)^{-1} v_\lambda(\boldsymbol{k}) 
\label{vlhf}
\end{equation}
starting from $v_0(\boldsymbol{k}) = \openone$. Note that $v_1(\boldsymbol{k})$ is analytic in $\boldsymbol{k}$ and satisfies the intrinsic symmetry $Xv_1(\boldsymbol{k})^*X=v_1(-\boldsymbol{k})$. According to Eq.~(\ref{vlhf}), this construction is also compatible with any additional symmetries.

\subsection{Strictly local and symmetric constructions for chiral symmetry classes}
\label{chiral}
In the previous subsection, we introduced a way to construct $q(\boldsymbol{k})$ out of $h_{\rm f}(\boldsymbol{k})$. Provided that $q(\boldsymbol{k})$ turns out to be local, it generally exhibits exponential tails even for strictly local $h_{\rm f}(\boldsymbol{k})$. Here we show that, for strictly local Hamiltonians in the chiral symmetry classes AIII, BDI and CII, which are known to be fully disentanglable \cite{Gong2021}, we can always construct strictly local and symmetric supercharges in any dimensions, no matter whether $h_{\rm f}(\boldsymbol{k})$ is topological or not.  

Let us start from class BDI, which respects a spinless time-reversal symmetry (TRS) alone: 
\begin{equation}
q(\boldsymbol{k})^*= q(-\boldsymbol{k}),\;\;\;\; h_{\rm f}(\boldsymbol{k})^* = h_{\rm f}(-\boldsymbol{k}).
\end{equation}
The general form of the supercharge is given by
\begin{equation}
q(\boldsymbol{k}) = \frac{\sigma_0 + \sigma_x}{2}\otimes q_+(\boldsymbol{k}) +  \frac{\sigma_0 - \sigma_x}{2}\otimes q_-(\boldsymbol{k}),
\label{BDISC}
\end{equation}
where $q_{\pm}(\boldsymbol{k})^*=q_{\pm}(-\boldsymbol{k})$, and that of the Hamiltonian reads
\begin{equation}
h_{\rm f}(\boldsymbol{k}) = \sigma_y\otimes h_y(\boldsymbol{k}) + \sigma_z\otimes h_z(\boldsymbol{k}),
\label{BDIH}
\end{equation}
where $h_{y,z}(\boldsymbol{k})^\dag=h_{y,z}(\boldsymbol{k})$, $h_y(\boldsymbol{k})^*= - h_y(-\boldsymbol{k})$ and $h_z(\boldsymbol{k})^*=h_z(-\boldsymbol{k})$. The supercharge (\ref{BDISC}) generates the following Hamiltonian:
\begin{equation}
\begin{split}
q(\boldsymbol{k})Z q(\boldsymbol{k})& = \sigma_y\otimes \frac{q_+(\boldsymbol{k})q_-(\boldsymbol{k})^\dag - q_-(\boldsymbol{k})q_+(\boldsymbol{k})^\dag}{2i} \\
&+ \sigma_z\otimes \frac{q_+(\boldsymbol{k})q_-(\boldsymbol{k})^\dag + q_-(\boldsymbol{k})q_+(\boldsymbol{k})^\dag}{2}.
\end{split}
\end{equation}
To make the above equation the target Hamiltonian (\ref{BDIH}), we should require:
\begin{equation}
\begin{split}
&q_+(\boldsymbol{k})q_-(\boldsymbol{k})^\dag - q_-(\boldsymbol{k})q_+(\boldsymbol{k})^\dag = 2i h_y(\boldsymbol{k}),\\
&q_+(\boldsymbol{k})q_-(\boldsymbol{k})^\dag + q_-(\boldsymbol{k})q_+(\boldsymbol{k})^\dag = 2h_z(\boldsymbol{k}),
\end{split}
\end{equation}
which admits the following symmetry-preserving (spinless TRS) and strictly local solution:
\begin{equation}
q_+(\boldsymbol{k})=\openone,\;\;\;\;q_-(\boldsymbol{k})= h_z(\boldsymbol{k}) - ih_y(\boldsymbol{k}).
\label{BDIloc}
\end{equation}
The strict locality follows from the fact that the entries of $q(\boldsymbol{k})$ are linear combinations of those of $h_{\rm f}(\boldsymbol{k})$, which is strictly local by assumption. 

We move on to class AIII, which respects a spin-$1/2$ TRS:
\begin{equation}
\begin{split}
&(\sigma_0\otimes\sigma_y\otimes\tilde\openone)q(\boldsymbol{k})^*(\sigma_0\otimes\sigma_y\otimes\tilde\openone)= q(-\boldsymbol{k}),\\
&(\sigma_0\otimes\sigma_y\otimes\tilde\openone)h_{\rm f}(\boldsymbol{k})^*(\sigma_0\otimes\sigma_y\otimes\tilde\openone) = h_{\rm f}(-\boldsymbol{k}),
\end{split}
\end{equation}
as well as a ${\rm U}(1)$ spin-rotation symmetry along $z$-axis:
\begin{equation}
[\sigma_z\otimes\sigma_z\otimes\tilde\openone, q(\boldsymbol{k})] = [\sigma_z\otimes\sigma_z\otimes\tilde\openone, h_{\rm f}(\boldsymbol{k})]=0.
\end{equation}
The general form of the supercharge is given by
\begin{equation}
q(\boldsymbol{k})=
\begin{bmatrix} 
q_1(\boldsymbol{k}) & 0 & 0 & q_2(\boldsymbol{k}) \\
0 & q_1(-\boldsymbol{k})^* & -q_2(-\boldsymbol{k})^* & 0 \\ 
0 & q_2(-\boldsymbol{k})^* & q_1(-\boldsymbol{k})^* & 0 \\ 
-q_2(\boldsymbol{k}) & 0 & 0 & q_1(\boldsymbol{k}) 
\end{bmatrix},
\label{AIIISC}
\end{equation}
where $q_1(\boldsymbol{k})$ and $q_2(\boldsymbol{k})$ are arbitrary (as long as $\det q(\boldsymbol{k})\neq0$), and that of the Hamiltonian reads
\begin{equation}
h_{\rm f}(\boldsymbol{k})=  
\begin{bmatrix} 
h_1(\boldsymbol{k}) & 0 & 0 & -h_2(\boldsymbol{k}) \\
0 & h_1(-\boldsymbol{k})^* & h_2(-\boldsymbol{k})^* & 0 \\ 
0 & h_2(-\boldsymbol{k})^* & -h_1(-\boldsymbol{k})^* & 0 \\ 
-h_2(\boldsymbol{k}) & 0 & 0 & -h_1(\boldsymbol{k}) 
\end{bmatrix},
\label{AIIIH}
\end{equation}
where $h_{1,2}(\boldsymbol{k})^\dag = h_{1,2}(\boldsymbol{k})$. The supercharge (\ref{AIIISC}) generates the following Hamiltonian:
\begin{widetext}
\begin{equation}
q(\boldsymbol{k})Z q(\boldsymbol{k})^\dag = 
\begin{bmatrix} 
(q_1q_1^\dag-q_2q_2^\dag)(\boldsymbol{k}) & 0 & 0 & -(q_1q_2^\dag+q_2q_1^\dag) (\boldsymbol{k}) \\
0 & (q_1^*q_1^{\rm T}-q_2^*q_2^{\rm T})(-\boldsymbol{k}) & (q_1^*q_2^{\rm T}+q_2^*q_1^{\rm T}) (-\boldsymbol{k}) & 0 \\ 
0 & (q_1^*q_2^{\rm T}+q_2^*q_1^{\rm T}) (-\boldsymbol{k}) & (q_2^*q_2^{\rm T}-q_1^*q_1^{\rm T})(-\boldsymbol{k}) & 0 \\ 
-(q_1q_2^\dag+q_2q_1^\dag) (\boldsymbol{k}) & 0 & 0 & (q_2q_2^\dag-q_1q_1^\dag)(\boldsymbol{k})
\end{bmatrix}.
\end{equation}
\end{widetext}
To make the above equation equal to the target Hamiltonian (\ref{AIIIH}), we should require:
\begin{equation}
\begin{split}
&q_1(\boldsymbol{k})q_1(\boldsymbol{k})^\dag-q_2(\boldsymbol{k})q_2^\dag(\boldsymbol{k})= h_1(\boldsymbol{k}),\\
&q_1(\boldsymbol{k})q_2(\boldsymbol{k})^\dag+q_2(\boldsymbol{k})q_1(\boldsymbol{k})^\dag= h_2(\boldsymbol{k}),
\end{split}
\end{equation}
which admits the following strictly local solution:
\begin{equation}
\begin{split}
q_1(\boldsymbol{k})&= \frac{1}{2}[h_1(\boldsymbol{k})-ih_2(\boldsymbol{k})+\openone],\\
q_2(\boldsymbol{k})&= \frac{i}{2}[h_1(\boldsymbol{k})-ih_2(\boldsymbol{k})-\openone].
\end{split}
\end{equation}

Finally, let us consider class CII. Ordering the internal degrees of freedom as Majorana, other and spin, the constraint of spin-$1/2$ TRS is given by
\begin{equation}
\begin{split}
&(\sigma_0\otimes\sigma_0\otimes\sigma_y\otimes\tilde\openone)q(\boldsymbol{k})^*(\sigma_0\otimes\sigma_0\otimes\sigma_y\otimes\tilde\openone)= q(-\boldsymbol{k}),\\
&(\sigma_0\otimes\sigma_0\otimes\sigma_y\otimes\tilde\openone)h_{\rm f}(\boldsymbol{k})^*(\sigma_0\otimes\sigma_0\otimes\sigma_y\otimes\tilde\openone) = h_{\rm f}(-\boldsymbol{k}),
\end{split}
\end{equation}
while that of the non-spin ${\rm SU}(2)$ symmetry reads
\begin{equation}
\begin{split}
&[\sigma_z\otimes\sigma_x\otimes \sigma_0\otimes\tilde\openone ,q(\boldsymbol{k})]=[\sigma_z\otimes\sigma_z \otimes \sigma_0 \otimes\tilde\openone ,q(\boldsymbol{k})]=0, \\
&[\sigma_z\otimes \sigma_x\otimes \sigma_0\otimes\tilde\openone ,h_{\rm f}(\boldsymbol{k})]=[\sigma_z\otimes \sigma_z\otimes \sigma_0\otimes\tilde\openone ,h_{\rm f}(\boldsymbol{k})]=0.
\end{split}
\end{equation}
The general form of the supercharge is thus given by
\begin{equation}
q(\boldsymbol{k})=\begin{bmatrix} 
\sigma_0\otimes \tilde a(\boldsymbol{k}) & \sigma_y\otimes \tilde b(\boldsymbol{k}) \\ 
\sigma_y\otimes \tilde b(-\boldsymbol{k})^* & -\sigma_0\otimes \tilde a(-\boldsymbol{k})^* 
\end{bmatrix},
\label{CIISC}
\end{equation}
where 
\begin{equation}
\begin{split}
&(\sigma_y\otimes\tilde\openone)\tilde a(\boldsymbol{k})^*(\sigma_y\otimes\tilde\openone)=\tilde a(-\boldsymbol{k}),\\
&(\sigma_y\otimes\tilde\openone)\tilde b(\boldsymbol{k})^*(\sigma_y\otimes\tilde\openone)=-\tilde b(-\boldsymbol{k}), 
\end{split}
\label{abTRS}
\end{equation}
and that of the Hamiltonian reads
\begin{equation}
h_{\rm f}(\boldsymbol{k})=  
\begin{bmatrix} 
\sigma_0\otimes \tilde A(\boldsymbol{k}) & \sigma_y\otimes \tilde B(\boldsymbol{k}) \\ 
\sigma_y\otimes \tilde B(-\boldsymbol{k})^* & -\sigma_0\otimes \tilde A(-\boldsymbol{k})^*
\end{bmatrix},
\label{CIIH}
\end{equation}
where $A(\boldsymbol{k})=A(\boldsymbol{k})^\dag$, $(\sigma_y\otimes\tilde\openone)A(\boldsymbol{k})^*(\sigma_y\otimes\tilde\openone)=A(-\boldsymbol{k})$ and $B(\boldsymbol{k})^*=B(-\boldsymbol{k})$, $(\sigma_y\otimes\tilde\openone)B(\boldsymbol{k})^*(\sigma_y\otimes\tilde\openone)=-B(-\boldsymbol{k})$. The supercharge (\ref{CIISC}) generates the following Hamiltonian:
\begin{widetext}
\begin{equation}
q(\boldsymbol{k})Z q(\boldsymbol{k})^\dag = 
\begin{bmatrix}
\sigma_0\otimes [\tilde a(\boldsymbol{k})\tilde a(\boldsymbol{k})^\dag - \tilde b(\boldsymbol{k})\tilde b(\boldsymbol{k})^\dag] & \sigma_y\otimes [\tilde a(\boldsymbol{k})\tilde b(-\boldsymbol{k})^{\rm T} + \tilde b(\boldsymbol{k})\tilde a(-\boldsymbol{k})^{\rm T}] \\
\sigma_y\otimes [\tilde b(-\boldsymbol{k})^*\tilde a(\boldsymbol{k})^\dag + \tilde a(-\boldsymbol{k})^*\tilde b(\boldsymbol{k})^\dag] &  \sigma_0\otimes [\tilde b(\boldsymbol{k})^*\tilde b(\boldsymbol{k})^{\rm T} - \tilde a(\boldsymbol{k})^*\tilde a(\boldsymbol{k})^{\rm T}]
\end{bmatrix}.
\end{equation}
\end{widetext}
To make the above equation equal to the target Hamiltonian (\ref{CIIH}), we should require:
\begin{equation}
\begin{split}
&\tilde a(\boldsymbol{k})\tilde a(\boldsymbol{k})^\dag - \tilde b(\boldsymbol{k})\tilde b(\boldsymbol{k})^\dag=\tilde A(\boldsymbol{k}),\\
&\tilde a(\boldsymbol{k})\tilde b(-\boldsymbol{k})^{\rm T} + \tilde b(\boldsymbol{k})\tilde a(-\boldsymbol{k})^{\rm T}=\tilde B(\boldsymbol{k}),
\end{split}
\end{equation}
which admits the following symmetry-preserving (spin-$1/2$ TRS) and strictly local solution:
\begin{equation}
\begin{split}
\tilde a(\boldsymbol{k}) &= \frac{1}{2}[\tilde A(\boldsymbol{k}) + \tilde B(\boldsymbol{k})(\sigma_y\otimes\tilde\openone) + \openone],\\
\tilde b(\boldsymbol{k}) &= \frac{1}{2}[\tilde A(\boldsymbol{k})(\sigma_y\otimes\tilde\openone) + \tilde B(\boldsymbol{k}) + (\sigma_y\otimes\tilde\openone)].
\end{split}
\end{equation}
One can check that Eq.~(\ref{abTRS}) is indeed fulfilled.

\subsection{Strictly local and asymmetric constructions for classes AI and AII}
\label{AIAII}
The general fermion Bloch Hamiltonian in class AI takes the following form: 
\begin{equation}
h_{\rm f}(\boldsymbol{k}) = \sigma_z\otimes h_{\rm I}(\boldsymbol{k}),\;\;\;\; h_{\rm I}(\boldsymbol{k})^*= h_{\rm I}(-\boldsymbol{k}).
\end{equation}
Note that class AI goes back to BDI if we forget the ${\rm U}(1)$ symmetry. Therefore, using the general construction for class BDI, we may choose the supercharge to be in the form of Eq.~(\ref{BDISC}), where
\begin{equation}
q_-(\boldsymbol{k})=\openone,\;\;\;\;q_+(\boldsymbol{k})=h_{\rm I}(\boldsymbol{k}).
\end{equation}
While this supercharge respects the TRS, the ${\rm U}(1)$ symmetry is broken in general. Moreover, whenever the Hamiltonian is strictly local, so is the supercharge.

For class AII, the general form of the fermion Bloch Hamiltonian reads 
\begin{equation}
h_{\rm f}(\boldsymbol{k}) = \frac{\sigma_0 + \sigma_z}{2}\otimes h_{\rm II}(\boldsymbol{k}) - \frac{\sigma_0 - \sigma_z}{2}\otimes h_{\rm II}(-\boldsymbol{k})^*, 
\label{HAII}
\end{equation}
where $(\sigma_y\otimes \tilde \openone)h_{\rm II}(\boldsymbol{k})^*(\sigma_y\otimes \tilde \openone) = h_{\rm II}(-\boldsymbol{k})$. Note that, similar to class AI, class AII goes back to DIII if we forget the ${\rm U}(1)$ symmetry. The general form of a supercharge in class DIII is given by
\begin{equation}
q(\boldsymbol{k})= \begin{bmatrix} 
a(\boldsymbol{k}) & b(\boldsymbol{k}) \\ 
b(-\boldsymbol{k})^* & a(-\boldsymbol{k})^*
\end{bmatrix},
\end{equation}
where $(\sigma_y\otimes \tilde \openone)a(\boldsymbol{k})^*(\sigma_y\otimes \tilde \openone) = a(-\boldsymbol{k})$ and $(\sigma_y\otimes \tilde \openone)b(\boldsymbol{k})^*(\sigma_y\otimes \tilde \openone) = b(-\boldsymbol{k})$. To generate Eq.~(\ref{HAII}), we only have impose
\begin{equation}
\begin{split}
&a(\boldsymbol{k})a(\boldsymbol{k})^\dag - b(\boldsymbol{k})b(\boldsymbol{k})^\dag = h_{\rm II}(\boldsymbol{k}),\\
&a(\boldsymbol{k})(\sigma_y\otimes\tilde\openone)b(\boldsymbol{k})^\dag = b(\boldsymbol{k})(\sigma_y\otimes\tilde\openone)a(\boldsymbol{k})^\dag,
\end{split}
\end{equation}
which can be readily satisfied by
\begin{equation}
a(\boldsymbol{k}) = \frac{\openone + h_{\rm II}(\boldsymbol{k})}{2},\;\;\;\;
b(\boldsymbol{k}) = \frac{\openone - h_{\rm II}(\boldsymbol{k})}{2}.
\end{equation}
Again, this supercharge is strictly local, provided that the Hamiltonian is.

\section{Full detail on Table~\ref{table1}}
\label{FDT1}
As mentioned in the main text, the problem concerning SUSY constructions can be related to the disentanglability of topological phases as discussed in Ref.~\cite{Gong2021}. Here we outline the derivations without going into technical calculations, and try to rephrase the argument in the language of SUSY whenever possible.

\subsection{Local and symmetric}
Let us first identify all the topological phases that can be generated by local and symmetric supercharges. The situation for the chiral-symmetry classes, including AIII, BDI and CII, is the simplest. As one has seen in the previous section, any local Hamiltonians in these classes can be generated by a local and symmetric supercharge. Indeed, one can even require the supercharge to be strictly local (i.e., with finite coupling range) whenever the Hamiltonian is strictly local. 

For the Wigner-Dyson classes, including A, AI and AII, one can straightforwardly compute the topological invariant for a Hamiltonian generated by a local and symmetric supercharge and always find a trivial result. This is because, after continuous (smooth) deformation by unitarizing $q(\boldsymbol{k})$, the obtained BdG Hamiltonian can be diagonalized by a unitary that is analytic and, for classes AI and AII, respects the TRS. This implies exponentially localized Wannier functions compatible with the TRS (if any), which are forbidden by a nontrivial topology.

Care should be taken for the remaining 4 BdG classes, including D, DIII, C and CI. If suffices to consider class D (C) in 0,1D (4,5D), and class DIII (CI) in 1,3,7D (3,5,7D), since otherwise $q(\boldsymbol{k})$ can be continuously trivialized into a constant while keeping the invertibility. For class D (C) in 0,1D (4,5D), there is a surjective homomorphism from the topological classes of BDI (CII) by forgetting the TRS. Since the latter can be generated by local and symmetric supercharges, so can the former. For class DIII (CI) in 3,7D, the topological number is given by the winding number. For a local and symmetric SUSY construction, the winding number is evaluated to be always even. This covers all the topological phases in 7D (3D), but only half in 3D (7D). Finally, for class DIII (CI) in 1D (5D), we have to compute the Chern-Simon form to determine the $\mathbb{Z}_2$ index. In the local and symmetric SUSY construction, where $q(\boldsymbol{k})$ is determined by a smaller matrix $a(\boldsymbol{k})$ with no symmetry constraint. The $\mathbb{Z}_2$ index turns out to be the parity of the winding number of $a(\boldsymbol{k})$ and thus can be nontrivial.

\subsection{Local but asymmetric}
As mentioned in the main text, to determine whether a topological phase can be generated by a local supercharge without requiring any symmetry constraint, it suffices to know its image in the classification group for class D under the inclusion map by forgetting all the symmetries. Then the answer is positive/negative, if the image can/cannot be generated by a local supercharge. 

Let us first consider the Wigner-Dyson classes. For classes AI and AII in $d=1,2,...,7$D, we can check that the image is always trivial, either from the triviality of the classification group for class D or that of the homomorphism from $\mathbb{Z}_2$ to $\mathbb{Z}$ or $2\mathbb{Z}$ \footnote{In fact, the homomorphism from any finite group to $\mathbb{Z}$ is always trivial.}. In 0D (more precisely, $d\equiv 0\mod 8$D), since the nontrivial $\mathbb{Z}_2$ topological phase in class D can be generated by a local supercharge, so can all the topological phases with additional symmetries. Applying the same argument to class A, we know that SUSY constructions based on local supercharges exist in 0D and 4D. Note that in the previous section, we have shown that all the local Hamiltonians in classes AI and AII can be generated by local supercharges by partially breaking the ${\rm U}(1)$ symmetry, which can be made strictly local whenever the Hamiltonians are. 
 
Regarding the BdG classes, we can check that the images of class DIII in 2,3D (CI in 6,7D) are trivial due to the very same reasons mentioned above. So far, we have identified all the topological phases that can be generated by local supercharges.
 
\subsection{Nonlocal}
Excluding the topological phases mentioned above, we finally single out the topological phases whose parent supercharges are necessarily non-local. These phases appear only in $d\equiv 2\mod 4$D in classes A, C and D, and are all characterized by nontrivial Chern numbers. 

Before ending the section, we would like to give another (on top of the Wannier localizaiton) physical interpretation to the absence of local parent supercharges for these chiral phases in 2D. It is well-known that in 2D the Hall conductance at zero temperature is proportional to the Chern number, and is thus nonzero if the system is in a topological phase. Suppose that the Hamiltonian can be generated by a local supercharge,  we can deform the Hamiltonian into a sum of (frustration-free) commuting local terms by unitarizing $q(\boldsymbol{k})$. Such a system has been proved to necessarily exhibit a vanishing Hall conductance \cite{Zhang2021}, leading to a contradiction.

\section{Particle-number conservation from gauge transformation}
\label{PCGT}
We prove that, given a local supercharge, it is always possible to gauge transform it to generate a ${\rm U}(1)$-symmetric (i.e., particle-number conserving or without pairing terms) SUSY boson Hamiltonian while keeping the fermion Hamiltonian fixed, and vice versa. 

\subsection{${\rm U}(1)$-symmetric $h_{\rm b}$ with fixed $h_{\rm f}$}
We first note that a gauge transformation in the supercharge $q(\boldsymbol{k})\to q(\boldsymbol{k})S(\boldsymbol{k})$, where $S(\boldsymbol{k})$ is a symplectic matrix satisfying 
\begin{equation}
S(\boldsymbol{k})Z S(\boldsymbol{k})^\dag = Z,\;\;\;\; XS(\boldsymbol{k})X = S(-\boldsymbol{k})^*,
\label{Sk}
\end{equation}
does not change $h_{\rm f}(\boldsymbol{k})$. On the other hand, $h_{\rm b}(\boldsymbol{k})= q(\boldsymbol{k})^\dag q(\boldsymbol{k})$ is turned into
\begin{equation}
h_{\rm b}(\boldsymbol{k}) = S(\boldsymbol{k})^\dag q(\boldsymbol{k})^\dag q(\boldsymbol{k}) S(\boldsymbol{k}).
\end{equation}
One can check that the spectrum of $Z h_{\rm b}(\boldsymbol{k})$ does not change since it undergoes a similarity transformation. To make $h_{\rm b}(\boldsymbol{k})$ particle-number conservation (i.e., without block-off-diagonal components), we have to impose the following constraint:
\begin{equation}
[Z, h_{\rm b}(\boldsymbol{k})]=0.
\label{Ghb}
\end{equation} 

Before finding $S(\boldsymbol{k})$ to validate Eq.~(\ref{Ghb}), we note that starting from any $S(\boldsymbol{k})$ satisfying Eq.~(\ref{Sk}), there always exists a unitary $U_S(\boldsymbol{k})$ and a Hermitian positive-definite matrix $|S(\boldsymbol{k})|$ determined from the polar decomposition $S(\boldsymbol{k})=U_S(\boldsymbol{k}) |S(\boldsymbol{k})|$, such that Eq.~(\ref{Sk}) is also satisfied. The right equation in Eq.~(\ref{Sk})  has been proved in Ref.~\cite{Gong2018}. To show the left equation, we can make use of the original relation for $S(\boldsymbol{k})$ to obtain 
\begin{equation}
\begin{split}
&S(\boldsymbol{k})^\dag S(\boldsymbol{k})Z S(\boldsymbol{k})^\dag =S(\boldsymbol{k})^\dag Z\;\;\Rightarrow\;\; \\
&S(\boldsymbol{k})^\dag S(\boldsymbol{k})Z S(\boldsymbol{k})^\dag S(\boldsymbol{k})Z S(\boldsymbol{k})^\dag S(\boldsymbol{k}) = S(\boldsymbol{k})^\dag S(\boldsymbol{k}) \\
&\Leftrightarrow\;\;Z|S(\boldsymbol{k})|^2Z=|S(\boldsymbol{k})|^{-2}.
\end{split}
\end{equation}
Recalling that $|S(\boldsymbol{k})|$ is positive-definite, the last equation above implies $Z|S(\boldsymbol{k})|Z=|S(\boldsymbol{k})|^{-1}$ \cite{Horn2013}. Substituting this relation into Eq.~(\ref{Sk}) gives
\begin{equation}
U_S(\boldsymbol{k})Z U_S(\boldsymbol{k})^\dag = Z\;\;\;\;\Leftrightarrow\;\;\;\; [U_S(\boldsymbol{k}),Z]=0,
\end{equation}
i.e., $U_S(\boldsymbol{k})$ is a particle-number-conserving operation. This fact further implies $S(\boldsymbol{k})\to U_S(\boldsymbol{k})^\dag S(\boldsymbol{k})$, which becomes Hermitian and positive definite, also validates Eqs.~(\ref{Sk}) and (\ref{Ghb}). Therefore, to construct number-conserving $h_{\rm b}(\boldsymbol{k})$, it suffices to consider Hermitian positive-definite symplectic matrices.

We claim that, under the above constraint (Eqs.~(\ref{Sk}) and (\ref{Ghb}) and $S(\boldsymbol{k})=S(\boldsymbol{k})^\dag>0$), the solution is uniquely given by
\begin{equation}
S(\boldsymbol{k})^2 = q(\boldsymbol{k})^{-1}|h_{\rm f}(\boldsymbol{k})| q(\boldsymbol{k})^{\dag -1},
\label{SqG}
\end{equation}
where we recall that $h_{\rm f}(\boldsymbol{k})= q(\boldsymbol{k})Zq(\boldsymbol{k})^\dag$. To confirm the symplecticity 
\begin{equation}
S(\boldsymbol{k})Z S(\boldsymbol{k})=Z, 
\label{SGS}
\end{equation}
it suffices to show $S(\boldsymbol{k})^2Z S(\boldsymbol{k})^2 = Z$, which is equivalent to
\begin{equation}
|h_{\rm f}(\boldsymbol{k})|h_{\rm f}(\boldsymbol{k})^{-1}|h_{\rm f}(\boldsymbol{k})|=h_{\rm f}(\boldsymbol{k}).\end{equation}
This is indeed true since $h_{\rm f}(\boldsymbol{k})$ is a positive-definite Hermitian matrix. To confirm Eq.~(\ref{Ghb}), which explicitly reads
\begin{equation}
S(\boldsymbol{k})|q(\boldsymbol{k})|^2S(\boldsymbol{k}) = Z S(\boldsymbol{k})|q(\boldsymbol{k})|^2S(\boldsymbol{k}) Z,
\end{equation}
it suffices to show
\begin{equation}
\begin{split}
&S(\boldsymbol{k})^2|q(\boldsymbol{k})|^2S(\boldsymbol{k})^2 \\
=&S(\boldsymbol{k}) Z S(\boldsymbol{k})|q(\boldsymbol{k})|^2S(\boldsymbol{k}) Z S(\boldsymbol{k}) \\
=&Z |q(\boldsymbol{k})|^2 Z,
\end{split}
\label{comb}
\end{equation}
where we have used the symplecticity (\ref{SGS}) of $S(\boldsymbol{k})$. This identity indeed holds true:
\begin{equation}
\begin{split}
&q(\boldsymbol{k})^{-1} \left|h_{\rm f}(\boldsymbol{k}) \right|^2  q(\boldsymbol{k})^{\dag-1} \\
=&q(\boldsymbol{k})^{-1} (q(\boldsymbol{k})Zq(\boldsymbol{k})^\dag)^2  q(\boldsymbol{k})^{\dag-1} \\
= &Z |q(\boldsymbol{k})|^2 Z.
\end{split}
\end{equation}
Finally, to see the uniqueness of $S(\boldsymbol{k})$, we only have to rewrite Eq.~(\ref{comb}) into
\begin{equation}
(q(\boldsymbol{k})S(\boldsymbol{k})^2q(\boldsymbol{k})^\dag)^2 = h_{\rm f}(\boldsymbol{k})^2. 
\end{equation}
Noting that both $q(\boldsymbol{k})S(\boldsymbol{k})^2q(\boldsymbol{k})^\dag$ and $\left|h_{\rm f}(\boldsymbol{k})\right|$ are positive-definite, the only solution should be Eq.~(\ref{SqG}) \cite{Horn2013}.

As a simple exercise, let us consider a two-band fermion system in class BDI and choose the original supercharge to be given by Eqs.~(\ref{BDISC}) and (\ref{BDIloc}) with $h_{y,z}(\boldsymbol{k})\in\mathbb{R}$.  After some straightforward calculations, one can obtain the gauge transformed supercharge $q(\boldsymbol{k})S(\boldsymbol{k})$ to be
\begin{equation}
\frac{\sqrt{\epsilon_{\boldsymbol{k}}}}{2}(\sigma_0+\sigma_x)+\frac{h_z(\boldsymbol{k})-ih_y(\boldsymbol{k})}{2\sqrt{\epsilon_{\boldsymbol{k}}}}(\sigma_0-\sigma_x),
\end{equation}
where $\epsilon_{\boldsymbol{k}}=\sqrt{h_y(\boldsymbol{k})^2 + h_z(\boldsymbol{k})^2}$. Note that the supercharge for the Kitaev chain in the main text follows this construction.

\subsection{${\rm U}(1)$-symmetric $h_{\rm f}$ with fixed $h_{\rm b}$}
We finally turn to the converse. For a fixed $h_{\rm b}(\boldsymbol{k})$, we have a gauge transformation $q(\boldsymbol{k})\to W(\boldsymbol{k}) q(\boldsymbol{k})$, where $W(\boldsymbol{k})$ is a unitary matrix satisfying
\begin{equation}
XW(\boldsymbol{k})^*X=W(-\boldsymbol{k}).
\end{equation}
To make the transformed fermion Hamiltonian particle-number conserving, i.e.,
\begin{equation}
\begin{split}
&[Z,W(\boldsymbol{k})h_{\rm f}(\boldsymbol{k}) W(\boldsymbol{k})^\dag]=0\;\;\;\;\Leftrightarrow\;\;\;\;\\
&[W(\boldsymbol{k})^\dag Z W(\boldsymbol{k}),h_{\rm f}(\boldsymbol{k})]=0,
\end{split}
\end{equation}
we can choose the involutory unitary $W(\boldsymbol{k})^\dag Z W(\boldsymbol{k})$ to be the flattened Hamiltonian (before transformation):
\begin{equation}
W(\boldsymbol{k})^\dag Z W(\boldsymbol{k}) = |h_{\rm f}(\boldsymbol{k})|^{-1} h_{\rm f}(\boldsymbol{k}).
\label{WZW}
\end{equation}
We claim that a possible solution is given by
\begin{equation}
W(\boldsymbol{k})= [q(\boldsymbol{k})^{-1}|h_{\rm f}(\boldsymbol{k})|q(\boldsymbol{k})^{\dag-1}]^{\frac{1}{2}}q(\boldsymbol{k})^\dag|h_{\rm f}(\boldsymbol{k})|^{-\frac{1}{2}}.
\end{equation}
First, let us check the unitarity:
\begin{equation}
\begin{split}
&W(\boldsymbol{k})W(\boldsymbol{k})^\dag = [q(\boldsymbol{k})^{-1}|h_{\rm f}(\boldsymbol{k})|q(\boldsymbol{k})^{\dag-1}]^{\frac{1}{2}} \\
&\times q(\boldsymbol{k})^\dag |h_{\rm f}(\boldsymbol{k})|^{-1}q(\boldsymbol{k}) [q(\boldsymbol{k})^{-1}|h_{\rm f}(\boldsymbol{k})|q(\boldsymbol{k})^{\dag-1}]^{\frac{1}{2}} \\
&=\openone.
\end{split}
\end{equation}
We then check Eq.~(\ref{WZW}), which turns out to be equivalent to
\begin{equation}
[q(\boldsymbol{k})^{-1}|h_{\rm f}(\boldsymbol{k})|q(\boldsymbol{k})^{\dag-1}]^{\frac{1}{2}}Z[q(\boldsymbol{k})^{-1}|h_{\rm f}(\boldsymbol{k})|q(\boldsymbol{k})^{\dag-1}]^{\frac{1}{2}}=Z.
\end{equation}
Since $q(\boldsymbol{k})^{-1}|h_{\rm f}(\boldsymbol{k})|q(\boldsymbol{k})^{\dag-1}$ is positive-definite, it suffices to show
\begin{equation}
q(\boldsymbol{k})^{-1}|h_{\rm f}(\boldsymbol{k})|q(\boldsymbol{k})^{\dag-1}Zq(\boldsymbol{k})^{-1}|h_{\rm f}(\boldsymbol{k})|q(\boldsymbol{k})^{\dag-1}=Z,
\end{equation}
which is nothing but $|h_{\rm f}(\boldsymbol{k})|h_{\rm f}(\boldsymbol{k})^{-1}|h_{\rm f}(\boldsymbol{k})|=h_{\rm f}(\boldsymbol{k})$ and thus indeed holds true.

\bibliography{GZP_references,library}
\end{document}